\documentclass[aps,pre,preprint,groupedaddress,showpacs]{revtex4-1}
\usepackage{amsmath,amsfonts,epsfig,graphics}
\usepackage{amssymb}
\usepackage{mathrsfs} 
\usepackage{amsbsy,color}
\usepackage{latexsym}
\usepackage{graphicx}

\pdfoutput=1   

\begin{document}

\hfill\ {DESY 18-090} 
\vskip1cm
\title{Testing the event-chain algorithm in asymptotically free models}

\author{Martin Hasenbusch}
\affiliation{Institut f\"ur Physik, Humboldt-Universit\"at zu Berlin, Newtonstr.~15, D-12489 Berlin, Germany}
\author{Stefan Schaefer}
\affiliation{John von Neumann Institute for Computing (NIC), DESY \\
    Platanenallee 6, D-15738 Zeuthen, Germany}

%
%
%
\begin{abstract}
We apply the event-chain algorithm proposed by Bernard, Krauth and Wilson
in 2009 to toy models of lattice QCD. We give a formal prove of stability 
of the algorithm. We study its performance at the example of the massive
Gaussian model on the square and the simple cubic lattice, 
the $O(3)$-invariant non-linear $\sigma$-model and the 
$SU(3) \times SU(3)$ principle chiral model on the square lattice. 
In all these cases we find that critical slowing down is essentially 
eliminated.
\end{abstract}


\maketitle

\section{Introduction}

Monte Carlo simulation of statistical field theory systems is a standard
tool in their study. While the  development of
algorithms has a long tradition, it is still an active field 
and source of substantial progress. Here we  
test a new class of recently proposed algorithms, the so-called
event-chain Monte Carlo~\cite{PhysRevE.80.056704,0295-5075-112-2-20003,PhysRevE.92.063306,KrauthPRL}, 
in systems with asymptotic freedom. 
In particular, we study the two-dimensional
$O(N)$ model and the chiral $SU(N)\times SU(N)$ principal 
chiral model for $N=3$.
As a preliminary step, we simulate free field theory.

The cost of a simulation is determined essentially by two factors: the numerical cost
of a single update step and the autocorrelation times $\tau_A$ of the observables $A$
of interest in units of these steps. The latter typically scale with
a power of the correlation length $\xi$:
\begin{equation}
\tau_A\propto \xi^z \,.
\end{equation}
The dynamical critical exponent $z$ therefore characterizes the algorithm's 
performance close to criticality. Algorithms, such as the local Metropolis
algorithm, which essentially perform
a random walk in configuration space have $z\approx 2$, whereas for some
models also update strategies whose autocorrelation times increase much slower
with the correlation length have been devised.
Examples are the cluster \cite{Swendsen:1987ce,Wolff:1988uh} and multigrid 
\cite{Goodman:1989jw} algorithms.
They are characterized by a coherent update of a large fraction of the 
field variables in one step.
Such advanced algorithms, however, rely on special features of the
theory to be simulated and are difficult to generalize: Algorithm performance
depends strongly on the model under investigation. For a new
proposal, it is therefore pivotal to test it in many cases to understand 
better its dynamics.

An interesting new entry in the toolbox is the so-called event-chain algorithm
pioneered in Ref.~\cite{PhysRevE.80.056704,0295-5075-112-2-20003,PhysRevE.92.063306}. 
It is basically a walker
on the lattice which updates the local field variable in Monte Carlo time
until a so-called event occurs, with probabilities given by the theory. Then
it moves to the neighboring site, which is determined by the event. 
The details are given below.
It is remarkable that for some theories this leads to a small value of $z$ and
not to the random walk one might naively expect.

\section{Event driven algorithm}
\label{eventalg}
In this section, the algorithm is discussed for a model with a single real 
variable 
at each site of the lattice. In order to apply the algorithm to more 
complicated models, the simple model is embedded, as we explain below.
Note that the idea of embedding is also used in the case of cluster
\cite{Brower:1989mt,Wolff:1988uh} and multigrid \cite{Mendesaetal96} algorithms.
To define the algorithm, one first discusses infinitesimal updates with
a step size $\epsilon > 0$. These updates are integrated analytically, 
leading to the version of the algorithm that is implemented at the end.

We consider a general scalar field theory with the action 
\begin{equation}
  S([\phi]) = \sum_{x\in \Lambda} s_x(\phi_x) + \sum_{\langle x,y\rangle}  s_{\langle x,y\rangle }(\phi_x,\phi_y) \;,
  \label{act}
\end{equation}
where  $s_x(\phi_x)$ and $s_{\langle x,y \rangle}(\phi_x,\phi_y)$ are 
functions of the real variables $\phi_x$.  Here we consider analytic 
functions. However the example of hard-spheres  considered in ref.
\cite{PhysRevE.80.056704} shows that this requirement can be relaxed.
The collection of all sites of the lattice 
is denoted by $\Lambda$ and $\langle x,y\rangle$ is a pair of  interacting
sites. In our numerical study, we consider nearest neighbor interactions only.
We require that 
\begin{equation}
\label{symab}
  \frac{\partial s_{\langle x,y\rangle}(\phi_x,\phi_y)}{\partial \phi_x} =
  -\frac{\partial s_{\langle x,y\rangle}(\phi_x,\phi_y)}{\partial \phi_y} 
\end{equation}
holds for the interaction terms.

The configuration space $\phi$ is now extended by two ``lifting'' variables:
the position of the walker $\tilde x\in \Lambda$ at which field updates are 
performed and the direction $\sigma\in \{-1,1\}$ of these changes.
Both are drawn from a uniform distribution, in particular uncorrelated to 
the fields $\phi$. 
 Let us denote an enlarged configuration by
\begin{equation}
 X=(\phi,\sigma, \tilde x)  \;.
\end{equation}

The action pertaining to a single site, keeping the fields at all other sites fixed,  can be written as
\begin{equation}
\label{singlesite}
  S_x(\phi_x)=s_{x}(\phi_x)+\sum_y s_{\langle x,y\rangle}(\phi_x,\phi_y) \equiv  \sum_{i=0}^n s_{i,x}(\phi_x) \, ,
\end{equation}
where $i=0$ indicates the single site term
and $i=1,2,...,n$ labels the interaction partners $y$ of $\tilde x$.

The enlarged configuration $X$ is always changed. Either
$\phi_{\tilde x}$ is replaced by $\phi_{\tilde x} + \sigma \epsilon$ or
$\sigma$ is replaced by $-\sigma$ or $\tilde x$ is replaced by one of its
interacting partners $y_i$.
The decisions are taken for each term of the single site 
action~(\ref{singlesite}) separately. 
With the standard Metropolis probability, applied to $s_{i,\tilde x}$ 
\begin{equation}
p_{i} = 
\mbox{min}\left[1,\exp[-s_{i,\tilde x}(\phi_{\tilde x}+\sigma \epsilon) 
+s_{i,\tilde x}(\phi_{\tilde x})] \right]
\end{equation}
the proposal $\phi_{\tilde x}+\sigma \epsilon$ is accepted.
Expanding in $\epsilon$ we get
\begin{equation}
\label{accepteps}
\begin{split}
p_{i} &= 
\begin{cases}
   1 - \sigma \frac{\partial s_{i,\tilde x}(\phi_{\tilde x})}{\partial \phi_{\tilde x}}
       \epsilon + O(\epsilon^2)
\;\;\; & \mbox{for}  \;\;\;
\sigma \frac{\partial s_{i,\tilde x}(\phi_{\tilde x})}{\partial \phi_{\tilde x}} > 0 \\
1 + O(\epsilon^2) \;\;\; & \mbox{for}   \;\;\;
\sigma \frac{\partial s_{i,\tilde x}(\phi_{\tilde x})}{\partial \phi_{\tilde x}}
\le  0 \;.
\end{cases}
\end{split}
\end{equation}
The update $\phi_{\tilde x}+\sigma \epsilon$ is rejected, if it is rejected
by at least one $s_{i,\tilde x}$. 
If the proposal is not accepted the following is done: In the case of 
$i=0$, $\sigma$ is replaced by $-\sigma$, while for $i>0$ the walker is set to
the site $y_i$. 
At finite $\epsilon$ there is a probability O$(\epsilon^2)$ that the 
proposal is rejected by more than one $s_{i,\tilde x}$. Hence for 
a sequence of $n$ updates with $t=n \epsilon$ finite, the  probability
for conflicting decisions is O$(\epsilon)$ and hence the algorithm 
becomes well defined in the limit $\epsilon \rightarrow 0$.
A proof of the correctness of the  algorithm is given in the Appendix.

The algorithm that is implemented in the program is obtained by integrating
infinitesimal steps over a finite fictitious time $t$. 
The probability that there is no rejection by $s_{i,\tilde x}$ in 
$n=t/\epsilon$ subsequent steps is
\begin{equation}
 P_{i}(\phi_{\tilde x}+\sigma t \leftarrow \phi_{\tilde x}) =
 \prod_{m=1}^n  p_i(\phi_{\tilde x} + m \sigma \epsilon \leftarrow 
\phi_{\tilde x} + (m-1) \sigma \epsilon) \;.
\end{equation}
Taking the logarithm and the limit $\epsilon \rightarrow 0$ one arrives at 
\begin{equation}
\label{Pfinite}
  \ln P_i(t) = \int_{0}^{t} \mbox{d} \psi \;\; 
  \mbox{min}[0,-\sigma s_{i,\tilde x}'(\phi_{\tilde x} + \sigma \psi)] 
=: - \Delta E_i(t) \;.
\end{equation}
Performing the integral we get
\begin{equation}
\Delta E_i(t) = \sigma
 \sum_k [s_{i,\tilde x}(\phi_{1,k}) - s_{i,\tilde x}(\phi_{0,k})] \;,
\end{equation}
where $k$ labels the intervals with $\sigma s_{i,\tilde x}'$ positive
throughout within $[\mbox{min}[\phi_{\tilde x}, \phi_{\tilde x} + \sigma t],
\mbox{max}[\phi_{\tilde x}, \phi_{\tilde x} + \sigma t]]$.  
$\phi_{0,k}$ and $\phi_{1,k}$ are the lower/upper and upper/lower bounds
of these intervals, depending on the value of  $\sigma$. 
The field $\phi_{\tilde x}$ is updated until the update is rejected by one of 
the $s_{i,\tilde x}$. 
The times $t_i^{(\mathrm{event})}$ when these events occur are determined
in the following way:
One draws a uniform random number $r_i\in(0,1]$ for each $i$ to fix $P_i(t)$. 
We arrive at Eq.~(11) of \cite{0295-5075-112-2-20003}:
\begin{equation}
\label{event}
 \Delta E_i \left(t_i^{(\mathrm{event})} \right) = - \ln r_i \,,
\end{equation}
which has to be solved for $t_i^{(\mathrm{event})}$.
The $i=i_{min}$ with the smallest $t_i^{(\mathrm{event})}$ makes the race:
$t^{(\mathrm{event})} =\mbox{min}_i[t_i^{(\mathrm{event})}]$.  The field 
is updated $\phi_{\tilde x} \rightarrow \phi_{\tilde x}
  + \sigma t^{(\mathrm{event})}$. For
$i_{min}=0$ we replace  $\sigma$ by $- \sigma$ and $\tilde x$ remains unchanged.
Else, for $i_{min}>0$ the walker assumes the new position $y_{i_{min}}$ and 
$\sigma$ keeps its sign.

The algorithm evolves in a fictitious time $t_{MC}$. For each event the 
fictitious time increases as $t_{MC} \rightarrow t_{MC} + t^{(\mathrm{event})}$.
A sequences of updates is started at $t_{MC}=0$ by drawing $\tilde x$ and 
$\sigma$ from a uniform distribution. The sequence of updates is stopped
at some fixed $t_f$. To this end, events are generated until $t_{MC}$ 
would overshoot $t_f$ for the first time. This last event is not taken into
account in the update of the field and $t_{MC}$ is not increased by
this last $t^{(\mathrm{event})}$. Instead 
$\phi_{\tilde x} \rightarrow \phi_{\tilde x} + \sigma (t_f - t_{MC})$
is performed. Furthermore, measurements should be performed in equal intervals
of the fictitious time. Mostly, we performed a measurement after a complete 
update 
sequences of the length $t_{MC}$ or after a fixed number of these sequences.
Note that measuring at events leads to a bias that, at least for small 
system sizes, can be easily seen in the averages of estimators.

In order to arrive at a formal proof of ergodicity one has to introduce
some randomness in the evolution time $t_f$. An alternative would be to 
add Metropolis updates that ensure ergodicity.  In our 
explorative study here, similar to refs. \cite{PhysRevE.80.056704,0295-5075-112-2-20003,PhysRevE.92.063306}, we simply ignore this question.

\section{The models}
We consider square or simple cubic lattices. 
The sites of the lattice are denoted by
$x=(x_0,x_1,...,x_{d-1})$ with $x_i \in \{0, 1, 2, ... L_i-1\}$, where 
$d$ is the dimension of the system,
setting the lattice spacing to unity, $a=1$. The directions of the lattice are
denoted by $\mu \in \{0,1,\dots,d-1\}$.
There are $2 d$ nearest neighbors $y_i= x \pm \hat \mu$, 
where $\hat \mu$ is a unit vector in $\mu$-direction.
In our simulations we take $L_0=L_1=...=L_{d-1} =L$ throughout.

\subsection{Free field theory}
First we study the two and three-dimensional scalar field theory.  It is defined by the
action
\begin{equation}
 S = \frac{1}{2} \sum_{x,\mu}  \left(\phi_x - \phi_{x + \hat \mu}   \right)^2  + \frac{m^2}{2} \sum_{x} \phi_x^2 + \frac{\lambda}{4!} \sum_{x} \phi_x^4
  \label{eq:sact}
\end{equation}
In the following we will only discuss the free case $\lambda=0$.

Here the functions~(\ref{Pfinite}) are easy to evaluate. For the single site 
term of the action we get 
\begin{equation}
  \begin{split}
    \Delta E_0(t)  &=  \frac{1}{2} m^2
    \begin{cases} 
        0  & \text{for} \qquad  \sigma \phi_{\tilde x}<0  \,, t < - \sigma \phi_{\tilde x} \\
  (\phi_{\tilde x} + \sigma t)^2 & \text{for} \qquad  \sigma \phi_{\tilde x} < 0  \,, t \ge - \sigma \phi_{\tilde x} \\
      (\phi_{\tilde x}+ \sigma t)^2 - \phi_{\tilde x}^2& \text{for} \qquad  \sigma \phi_{\tilde x} \ge 0\, .
    \end{cases}
  \end{split}
\end{equation}
For $i>0$ we get
\begin{equation}
 \begin{split}
  \Delta  E_i(t)  &=  \frac{1}{2} 
    \begin{cases}
        0  & \text{for} \qquad  \sigma \Delta \phi_{\tilde x,i}<0  \,, t < - \sigma \Delta \phi_{\tilde x,i} \\
  (\Delta \phi_{\tilde x,i} + \sigma t)^2 & \text{for} \qquad  
\sigma \Delta \phi_{\tilde x,i} < 0  \,, t \ge - \sigma \Delta \phi_{\tilde x,i} \\
      (\Delta \phi_{\tilde x,i}+ \sigma t)^2 - \Delta \phi_{\tilde x,i}^2& \text{for} \qquad  \sigma \Delta \phi_{\tilde x,i} \ge 0\, ,
    \end{cases}
  \end{split}
\end{equation}
where $\Delta \phi_{\tilde x,i} =  \phi_{\tilde x} -  \phi_{\tilde x+ \hat \mu}
 =  \phi_{\tilde x} -  \phi_{y_i}$. 
There is always a solution for a positive $t$. 

\subsection{Non-linear $\sigma$-model}
The action of the non-linear $O(N)$-invariant $\sigma$-model is
\begin{equation}
 S = - \beta \sum_{x, \mu} \vec{s}_x \cdot \vec{s}_{x+\hat \mu}   \;,
  \label{eq:oact}
\end{equation}
where $\vec{s}_x$ is a unit vector with $N$ real components. We shall study the
model for $N=2$ and $3$. For $N=2$ we get the so called XY-model.
It undergoes a Kosterlitz-Thouless phase transition at  
$\beta_{KT}= 1.1199(1)$ \cite{HaPi96}.
For temperatures above the transition, there is a finite
correlation length. At lower temperatures there is no ordering and
the two-point correlation function is decaying with a power law.  
For references see for example \cite{Ha08}.
For the event-chain algorithm it is useful to write the XY-model in terms of 
angles $\alpha_x$:
\begin{equation}
 S = - \beta \sum_{x, \mu} \cos(\alpha_x - \alpha_{x+\hat \mu})   \;.
  \label{eq:oact2}
\end{equation}
The pair terms of the action are
\begin{equation}
s_{i,\tilde x} = - \beta \cos(\alpha_{\tilde x} - \alpha_{y_i}) \;.
\end{equation}
In the update,  $\alpha_{\tilde x}$ is incremented.
The functions~(\ref{Pfinite}) are given in ref. \cite{0295-5075-112-2-20003}. To simplify
the notation, let us assume $\sigma = 1$ in the following. Let us define
\begin{equation}
\delta_i = \alpha_{\tilde x} - \alpha_{y_i}  - 2 m \pi \;,
\end{equation}
where the integer $m$ is chosen such that $- \pi < \delta_i \le \pi $. 
\begin{equation}
\label{XYenergy}
 \begin{split}
  \Delta  E_i(t)  &=   \beta
    \begin{cases}
2 n& \text{for} \qquad \delta_i <0 \,, \delta_i + t - 2 n \pi < 0  \, ,\\
2 n + 1 -\cos(\delta_i+t) & \text{for} \qquad \delta_i <0 \,, \delta_i + t - 2 n \pi \ge 0 \, ,  \\
2 (n-1) + \cos(\delta_i) -1 & \text{for} \qquad \delta_i \ge 0 \,, \delta_i + t - 2 n \pi < 0  \, , \\
2 n + \cos(\delta_i)  -\cos(\delta_i+t) & \text{for} \qquad \delta_i \ge 0 \,, \delta_i + t- 2 n \pi \ge 0 \, , 
\end{cases}
\end{split}
\end{equation}
where the integer $n \ge 0$ is chosen such that 
$- \pi <  \delta_i + t - 2 n \pi \le \pi $.   Note that  $2 n$ is
the contribution from $n$ complete cycles, where $\cos(0) - \cos(\pi) = 2$.
For an illustration see Fig.~4 of ref. \cite{0295-5075-112-2-20003}.

For $N=3$ the model has a finite correlation length at any finite value
of $\beta$.  The model is asymptotically free. The divergence of the correlation
length as $\beta \rightarrow \infty$ is governed by the so called
$\beta$-function.  For a discussion of the physics of this model see 
for example \cite{Balog:1999ww}. 

In the literature there are a number of Monte Carlo studies of this
lattice model. It can be efficiently simulated by using the cluster
algorithm \cite{UWolff89} and the multigrid algorithm \cite{Mendesaetal96}. 
Also the micro-canonical
overrelaxation algorithm \cite{Apo91} had been applied successfully.
For simulations with the worm algorithm see ref. \cite{UlliWurm}. 
The simulation for $N>2$ with the event-chain Monte Carlo algorithm 
is performed, by an embedding of a generalized XY model. To this end, 
for one sequence of updates, one picks out a pair $(l,k)$ of components 
from the 
$N$ components of the field. In the update, only rotations in this plane are
performed. The algorithm is run as for the XY model, with 
the exception that in eq.~(\ref{XYenergy}) the prefactor $\beta$ is 
replaced by $\sqrt{s_{l,\tilde x}^2 s_{k,\tilde x}^2} 
\sqrt{s_{l,y_i}^2 s_{k,y_i}^2} \beta$.

\subsection{$SU(N)\times SU(N)$ principal chiral model}
\label{principalchiral}
The principal chiral model on the square lattice is defined by the action
\begin{equation}
  S =-\beta \sum_{x,\mu} \mathrm{Re}\, \mathrm{tr}\, U_x U_{x+\hat \mu} 
\end{equation}
where $U_x \in $SU$(N)$. 
The action restricted to a single site is 
\begin{equation}
  S_x=-\beta  \sum_{\pm \hat \mu}\, \mathrm{Re} \, \mathrm{tr}\, U_x U_{x+\hat\mu}^\dagger\equiv \sum_{i=1}^{2d} s_{x,i}\,.
\end{equation}
The model is asymptotically free. For a discussion of the physics of this 
model see for example \cite{Mana:1996pk}. No efficient implementation of 
the cluster algorithm has been devised for this model so far.  However 
the multigrid algorithm has been implemented successfully 
\cite{Mana:1996pk}. To this end, for one sequence of updates a  
one-parameter subgroup is considered.  Here we apply the same idea.
Following \cite{Mana:1996pk}, a general one parameter update can be 
composed of a random SU$(N)$ rotation and a
rotation by a variable angle along one, fixed element $\lambda$ of the 
algebra. With 
\[
\lambda=\begin{pmatrix} i & 0 & \dots\\ 0 & -i & 0 & \dots\\ 0 & \dots&\\  \end{pmatrix}
\]
the updates read
\[
  U \to e^{t R \lambda R^\dagger} U = R \; e^{t \lambda}  R^\dagger U \,,
  \label{eq:sunup}
\]
leading to an embedded action for site $x$ and component $i$
\[
 s _{i,x}(t) = -\beta [a \cos t + b \sin t] = - \beta c  \cos(t - t_0) 
\]
with
\begin{equation}
    a = \mathrm{Re}([W_i]_{11}+[W_i]_{22}) \qquad  \mbox{and} \qquad b= \mathrm{Im}([W_i]_{11}- [W_i]_{22})  \; ,
\end{equation}
where  $W_i=R^\dagger U_{x} U^\dagger_{y_i} R$.  Furthermore
$c = \sqrt{a^2+b^2}$ and $t_0 = \arctan(b/a)$. 

\section{Numerical results}
Below we discuss our numerical results for the scalar free field theory, 
the $O(3)$-invariant nonlinear $\sigma$-model and the 
$SU(3) \times SU(3)$-invariant principal chiral model. We abstain 
from a discussion of our simulations of the two-dimensional XY-model, 
since our results are fully consistent with those of ref. \cite{Lei:2018vis}. 

\subsection{Scalar free field theory}
We simulated the Gaussian model on the square and simple cubic lattice.
In our simulations, we have taken $L \propto m^{-1}$, 
evolving the fictitious Monte Carlo time for the period $t_f$ using 
the event-chain algorithm. Once this is reached, a new site $\tilde x$ and 
direction $\sigma$ are selected. Note that  $t_f$ is the only free parameter
of the algorithm.

Let us define the quantities that we measured in  our simulations.
The first two quantities are related to the terms in the action:
\begin{align}
  O_1 &= \frac{1}{2 d L^d} \sum_{<xy>} (\phi_x - \phi_y)^2 \,\, , & 
  O_2 &= \frac{1}{2 L^d} \phi_x^2 \;.
\end{align}
Then we determined magnetization and a proxy of the correlation length  
\begin{align}
  O_3 &= \frac{1}{2}  \left( \sum_x \phi_x \right)^2 \,\, ,&
  O_4 &= \sum_{i=0}^{d-1}  \tilde \phi_i  \tilde \phi_i^*
\end{align}
with the Fourier transform $ \tilde \phi_i = \sum_x \exp(-i 2 \pi x_i/L) \phi_x$.

In the Gaussian model, all these quantities can be easily computed 
exactly by Fourier transformation.
We compared the outcome of our simulations with these exact results. 
The largest deviation had been  $3.4$ standard deviations for a single 
observable.

We performed a measurement after $n_\mathrm{m}$ sequences of the length $t_f$. 
For a given $L$ we kept $t_\mathrm{meas} = n_\mathrm{m} t_f$ constant. It is chosen 
such that the integrated autocorrelation times in units of 
$t_\mathrm{meas}$ are of $O(10)$.  

\subsubsection{Two dimensions}
In the two-dimensional 
case we simulated the linear lattice sizes $L=32$, $64$, $128$, and $256$.
Correspondingly the masses are taken as $m=0.3$, $0.15$, $0.075$, and
$0.0375$, respectively.  Each  series of simulations  consists of 
$10^6$ measurements, separated by   $t_\mathrm{meas}=100$, $400$, $2000$ and 
$8000$ for $L=32$, $64$, $128$, and $256$, respectively. 
We focus on the dependence of the autocorrelations
on the length $t_f$ of the update sequence.

\begin{figure}
\begin{center}
\includegraphics[width=0.45\textwidth]{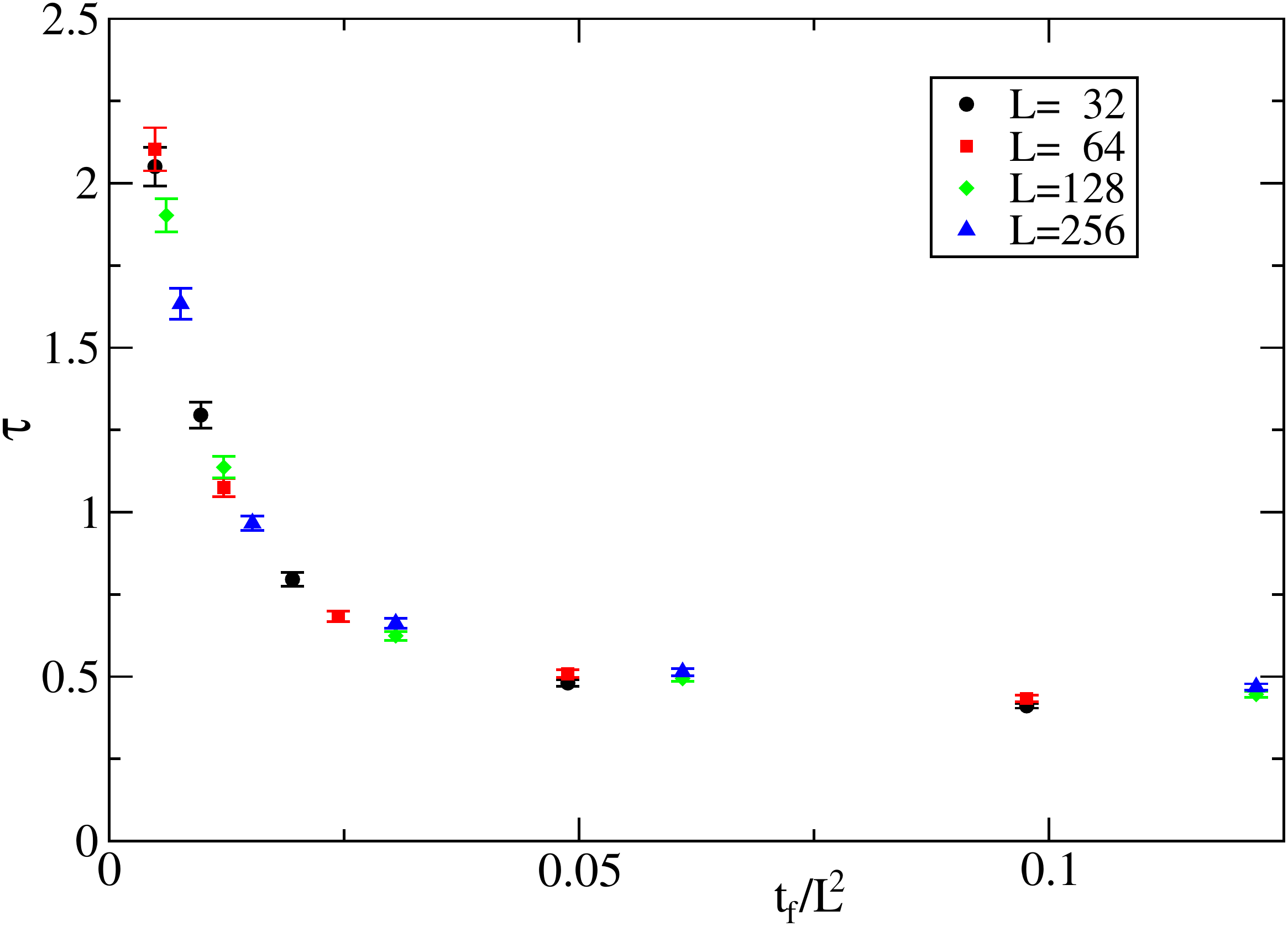}
  \includegraphics[width=0.45\textwidth]{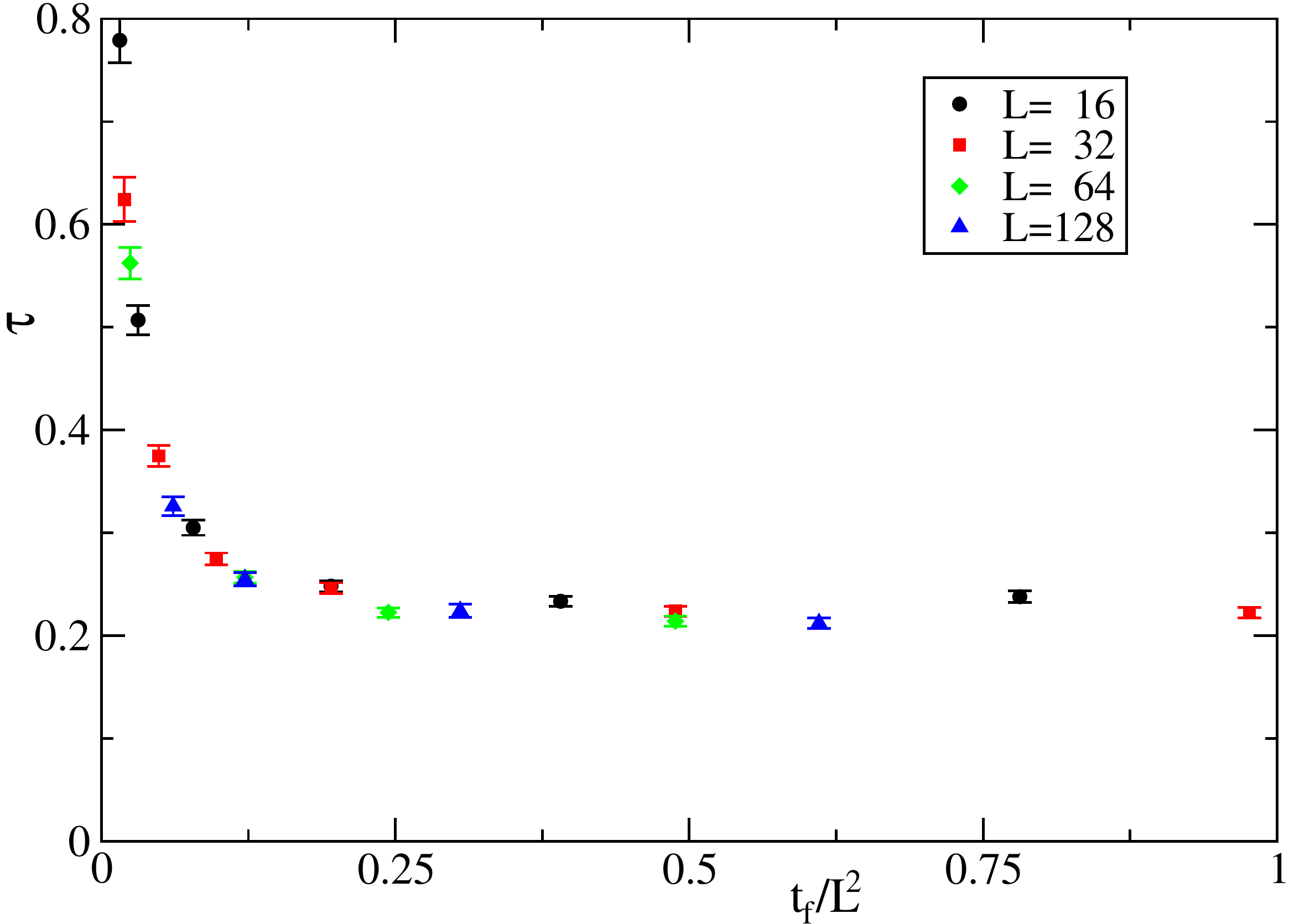}
\caption{\label{tau2DO3} Scaled integrated autocorrelation time for the 
observable $O_3$, left the two-dimensional Gaussian model, right the 
three-dimensional version.}
\end{center}
\end{figure}

In Fig.~\ref{tau2DO3} we give the integrated autocorrelation time of $O_3$, 
while for the other observables the results are listed in table 
\ref{tau2Dtable}.
Since the cost of the simulation scales with $L^2$ at least, 
the autocorrelation times are given in units of $L^2$.  This means that the
autocorrelation times in units of measurements are divided  by $32^2/100$,
$64^2/400$, $128^2/2000$,  $256^2/8000$, respectively. Plotting the 
autocorrelation time as  a function of $t_f/L^2$, we see a reasonable collapse of the
data for the four different lattice sizes. We conclude that $t_f$ should  be
chosen such that $t_f \propto m^{-2} \propto L^2$.   The integrated
autocorrelation time of $O_3$ decreases with increasing $t_f$ and seems to
approach a plateau value for large $t_f$.  
We can not exclude that
$\tau$ increases for very large $t_f$ again, we will see an example below in the 
principle chiral model.  Here we will not examine
$t_f > L^2$.

From the plot
we see a slight increase of $\tau$ with increasing $L$. This will be studied
below for a fixed value of $t_f/L^2$ using higher statistics.  The behavior of
$\tau_{int}$  for $O_4$ is similar to that of $O_3$.  In the case of $O_1$ and
$O_2$ the behavior is different.  For  $O_1$ we see an increase of
$\tau_{int}$ with increasing $t_f$.  This increase is not dramatic and
$\tau_{int}$ seems to approach a plateau. For $O_2$ we see a decrease of
$\tau_{int}$ with increasing $t_f$. It is however less pronounced as for $O_3$
and $O_4$.

Next we performed simulations for $t_f/L^2=125/4096\approx 0.030517578...$
fixed.  Measurements were performed after $t_\mathrm{meas}= 2 t_f$.   Here 
we performed $5 \times 10^6$ measurements for each lattice size.  In addition
to  $L=32$, $64$, $128$, and $256$ we have simulated $L=512$.  The 
simulation for $L=512$ took 8 hours on one core of a 
Intel(R) Xeon(R) CPU E5-2660 v3  2.60GHz. The results for the 
integrated autocorrelation times are summarized in table \ref{tau2Dtable}.

\begin{table}
\caption{\sl \label{tau2Dtable}
Integrated autocorrelation times for $t_f/L^2=125/4096\approx 0.030517578...$.
The distance between two subsequent measurements is $t_\mathrm{meas} =  2 t_f$. 
For each $L$, $5 \times 10^6$ measurements are performed. The $\tau$ are 
given in units of $t_\mathrm{meas}$.
}
\begin{center}
\begin{tabular}{rrrrr}
\hline
 $L$& $\tau_{\mathrm{int},1}$ & $\tau_{\mathrm{int},2}$ & $\tau_{\mathrm{int},3}$ &  $\tau_{\mathrm{int},4}$\\
\hline
  32 &  7.23(5) &  4.80(3) & 3.70(3)& 3.56(2) \\
  64 &  8.03(5) &  4.80(3) & 3.90(4)& 3.79(3) \\
 128 &  8.74(5) &  4.95(3) & 4.10(4)& 3.99(3) \\
 256 &  9.47(6) &  5.09(3) & 4.26(4)& 4.18(3) \\
 512 & 10.13(6) &  5.16(3) & 4.42(4)& 4.39(3) \\
\hline
\end{tabular}
\end{center}
\end{table}

Looking at the numbers we see that for each observable, the integrated
autocorrelation time is increasing with increasing lattice size. Roughly, 
doubling the lattice size, the autocorrelation times increase by an 
additive constant. I.e. its seems that the autocorrelation times 
increase logarithmically in the lattice size.

The CPU time that is used is proportional to the number of 
events. There is a small dependence on the mass and the lattice size
for the number of events divided by $t_f$. Roughly this ratio 
equals $1.128$ for all our masses and lattice sizes.

Note that our results are consistent with those of ref. \cite{Lei:2018vis}, 
where the massless case was studied.

\subsubsection{Three dimensions}
Next we simulated the three-dimensional Gaussian model.  
We simulated the linear lattice sizes $L=16$, $32$, $64$, and
$128$ for various values of $t_f$. Similar to the two-dimensional case,  
we use the masses $m=0.6$, $0.3$, $0.15$, and $0.075$.
In Fig.~\ref{tau2DO3} on the right, we give the integrated autocorrelation 
time of the observable $O_3$.
Also in three dimensions it turns out that the curves for the different 
$L$ fall approximately on top of each other for plotting $\tau$ as a 
function of $t_f/L^2$.  Since we have taken $L \propto m^{-1}$, it is likely, 
that actually $m^2 t_f$ should be kept constant.  
\begin{table}
\caption{\label{tau3Dtable}
Autocorrelation times for the three-dimensional Gaussian model.
Integrated autocorrelation times for $t_f/L^2=1/4 =0.25$.
The distance between two subsequent measurements is $t_\mathrm{meas} = L^3/32 = 
t_f L/8$.
For each $L$, $5 \times 10^6$ measurements are performed. The $\tau$ are
given in units of $t_\mathrm{meas}$.
}
\begin{center}
\begin{tabular}{rrrrr}
\hline
  $L$& $\tau_{\mathrm{int},1}$ & $\tau_{\mathrm{int},2}$ & $\tau_{\mathrm{int},3}$ &  $\tau_{\mathrm{int},4}$\\
\hline
  16 & 17.32(15) & 13.00(11) & 7.85(9) & 7.59(5) \\
  32 & 16.71(14) & 10.53(9)  & 7.56(8) & 7.12(5) \\
  64 & 17.08(14) &  9.02(8)  & 7.45(8) & 7.04(5) \\
 128 & 16.82(14) &  7.75(5)  & 7.29(7) & 6.87(5) \\
\hline
\end{tabular}
\end{center}
\end{table}

As for two dimensions, we have measured the number of events per $t$.
We get here  $\approx 1.38$. 
For $t_f/L^2=1/4 =0.25$ we performed simulations with $5 \times 10^6$ 
measurements.  The results of these simulations are summarized in 
table \ref{tau3Dtable}.  In contrast to two dimensions, the autocorrelation 
times are slightly decreasing with increasing lattice size (decreasing mass). 

\subsection{The $O(3)$-invariant nonlinear $\sigma$-model in two dimensions}
We performed updates in the $(0,1)$, $(0,2)$ and $(1,2)$ planes
of the spins in a fixed order.
For each of the planes, we used $t_f = L_0 L_1$.
We simulated the model at $\beta=1.4, 1.5, 1.6, 1.7$,  and $1.8$ on lattices of the linear size
$L=68, 110, 190, 346$, and $646$, respectively.  Following ref. \cite{UWolff89}  
the correlation length is $\xi=6.90(1)$, $11.09(2)$, $19.07(6)$, $34.57(7)$ and $64.78(15)$
at these values of $\beta$, respectively. Consistent results are given in 
ref. \cite{balogetal99}. 

Our results for the observables 
\begin{equation}
E = \frac{1}{L^2} \sum_{x,\mu}  \vec{s}_x \cdot \vec{s}_{x +\hat \mu}  
\, ,  \;\;\,  \chi=\frac{1}{L^2} \left(\sum_x \vec{s}_x \right)^2
\end{equation}
and the corresponding integrated autocorrelation times
are summarized in table \ref{aaa}. In the last column we give the average number of events divided
by the number of sites for one cycle of three XY embeddings. The number slowly increases with
increasing lattice size. This has to be taken into account when judging the performance of the
algorithm. The fact that the number is close to one in all cases means that in one cycle, each site
is touched roughly once.
\begin{table}
\caption{\sl \label{aaa}  
Our results for the $O(3)$-invariant nonlinear $\sigma$-model on the
square lattice.   
}
\begin{center}
\begin{tabular}{cllllll}
\hline
$\beta$  & $E$ &  $\tau_E$ & $\chi$ & $\tau_{\chi}$ &  $N_{ev}/(L_0 L_1)$  \\
\hline
  1.4 & 1.124340(23)  & 1.476(10)  & 78.65(10)  & 1.358(9)  & 0.954050(12) \\
  1.5 & 1.203250(14)  & 1.507(11)  &176.67(25)  & 1.597(13) & 0.995291(9) \\
  1.6 & 1.271421(7)   & 1.503(12)  &447.42(69)  & 1.937(17) & 1.033550(6)  \\
  1.7 & 1.3284761(35) & 1.376(12)  &1273.9(2.1) & 2.306(26) & 1.070078(4)  \\
  1.8 & 1.3758734(16) & 1.256(11)  & 3841.2(7.0)& 2.739(31) & 1.105618(3) \\
\hline
\end{tabular}
\end{center}
\end{table}

In the case of the energy density $E$, the integrated autocorrelation time 
even decreases with increasing correlation length $\xi$. On the other hand, 
for the susceptibility $\chi$, the integrated autocorrelation time is 
increasing with increasing $\xi$. Similar to the two-dimensional free 
field case, this increase seems to be logarithmic in $\xi$.  Note that 
in ref. \cite{PhysRevE.92.063306}  slowing down with $z \approx 1$ has been  
observed for 
the $O(3)$-invariant model on the three-dimensional simple cubic lattice.
Furthermore one should notice that the single cluster algorithm \cite{UWolff89}
is more efficient than the event-chain Monte Carlo algorithm 
and on top of that provides variance reduced estimators for various 
observables.

\subsection{The $SU(3) \times SU(3)$-invariant principal chiral model}
Following Ref.~\cite{Hasenbusch:1991aj,Mana:1996pk}, we have investigated the performance of the event driven 
algorithm also in the $SU(3) \times SU(3)$-invariant principal chiral model. 
An update sequence is defined
as a fixed random rotation matrix $R$, which is used in the event algorithm until a total
rotation $t_f$ has been performed. Then for the next update 
sequence a new $R$ is used. 
In table \ref{t:1} we give the 
basic parameters of our runs.  Again we confirm that the fluctuations of the results from various
runs and trajectory lengths are compatible with the observed statistical noise and agree also
with results from the literature \cite{Hasenbusch:1991aj}.

We will focus on the following observables, constructed from fundamental and
adjoint correlation functions \cite{Mana:1996pk}
\begin{equation}
  G_F(x-y)=\langle \mathrm{tr}\,(U_x^\dagger U_y)\rangle \quad  ; \qquad \qquad G_A(x-y)=\langle|\mathrm{tr} ( U_x^\dagger U_y)|^2\rangle -1\,.
\end{equation}
using 
  \begin{align}
    E_F&=\frac{1}{N} G_F(1,0) ;&  E_A&=\frac{1}{N^2-1} G_A(1,0)\\
    \chi_F&=\sum_{x\in\Lambda} G_F(x) ;&  \chi_A&=\sum_{x\in\Lambda} G_A(x)
  \end{align}
  as well as the correlation length
  \begin{align}
    \xi_F&=\frac{(\chi_F/F_F-1)^{1/2}}{2\sin(\pi/L)}\ ;& F_F=\sum_x e^{2\pi i x_0/L}G_F(x)\,.
  \end{align}

\begin{table}
  \caption{Parameters of the runs for the principle chiral model.\label{t:1}}
  \begin{center}
    \begin{tabular}{ccccc}
\hline
      $\beta$ & $L$ & $\xi_F$ & $L/\xi_F$ \\
\hline
      1.5     &  32       &  3.01629(4)& 10.6 \\
      1.65    &  64       &  5.4549(2) & 11.8 \\
      1.825   &  128      &  11.679(6) & 11.0 \\
      1.985   &  256      &  23.41(5)  & 10.9 \\
      \hline
    \end{tabular}
  \end{center}
\end{table}

\begin{figure}
  \begin{center}
    \includegraphics[width=0.48\textwidth]{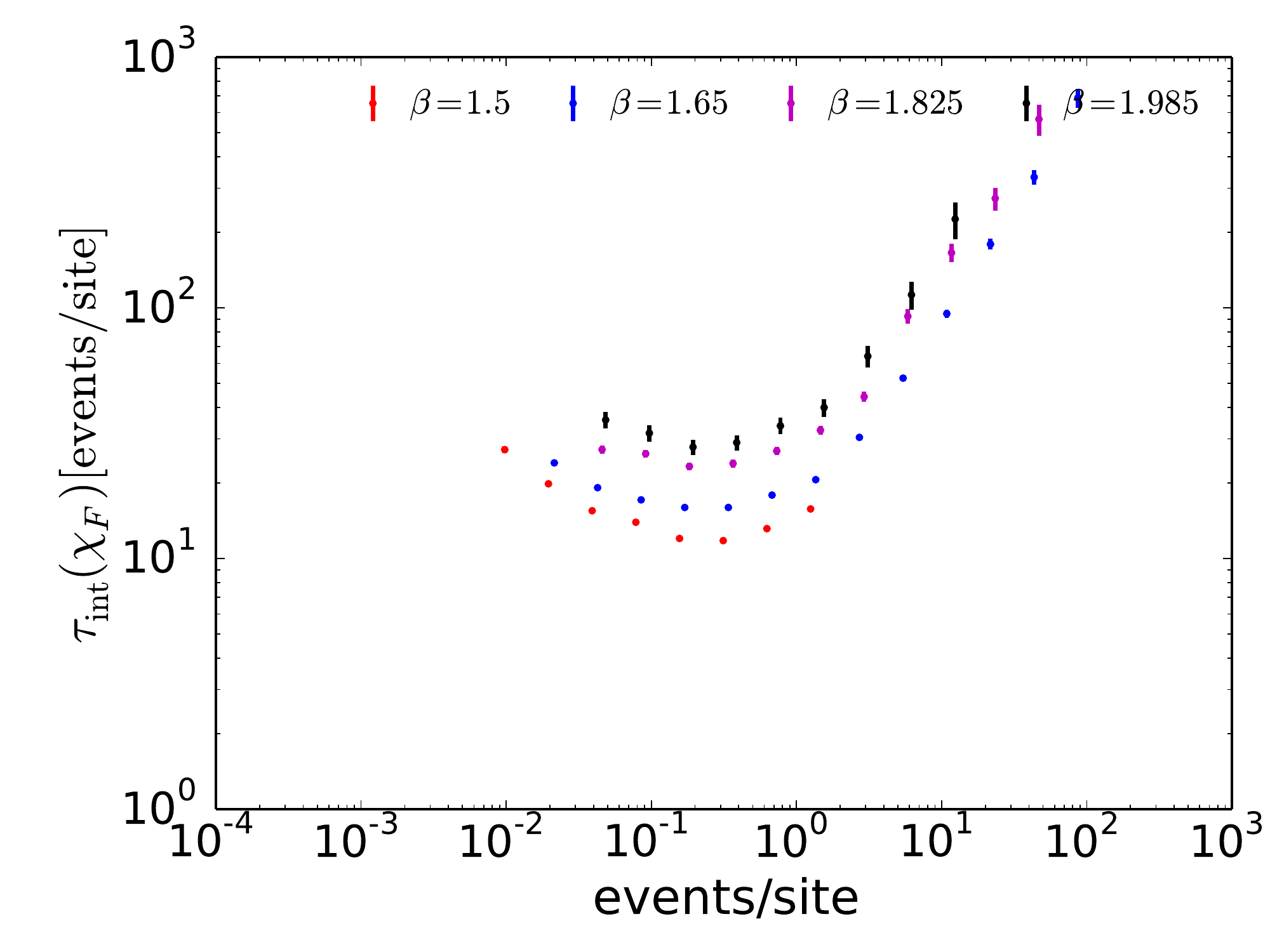}
    \includegraphics[width=0.48\textwidth]{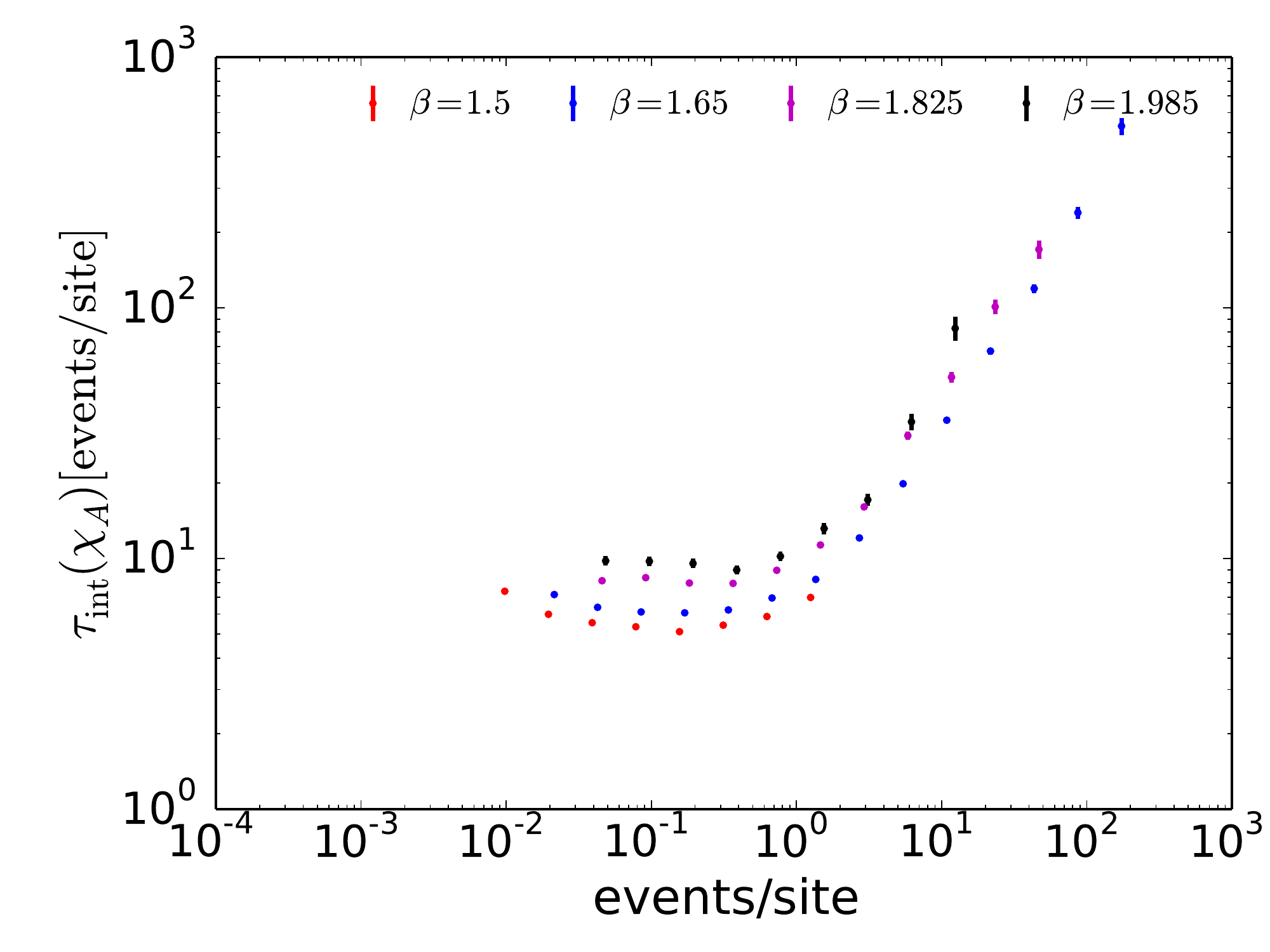}
  \end{center}
  \caption{Integrated autocorrelation time of the susceptibility as a function
   of the number of events within one update sequence divided 
  by the lattice volume. One the left/right we give the results
 for the fundamental/adjoint representation.
  In both cases, we observe a weak dependence and a shallow minimum. \label{f2}}
\end{figure}

In this particular study, we also measure in time intervals which are smaller than  $t_f$
in order to reveal the scaling behavior with long update sequences.
The results are  shown in Fig.~\ref{f2}. 
We observe a rather weak dependence, with a shallow minimum, again for 
sequences with a length such that $0.1$ to $1$ events per site occur.

\begin{figure}
  \begin{center}
    \includegraphics[width=0.50\textwidth]{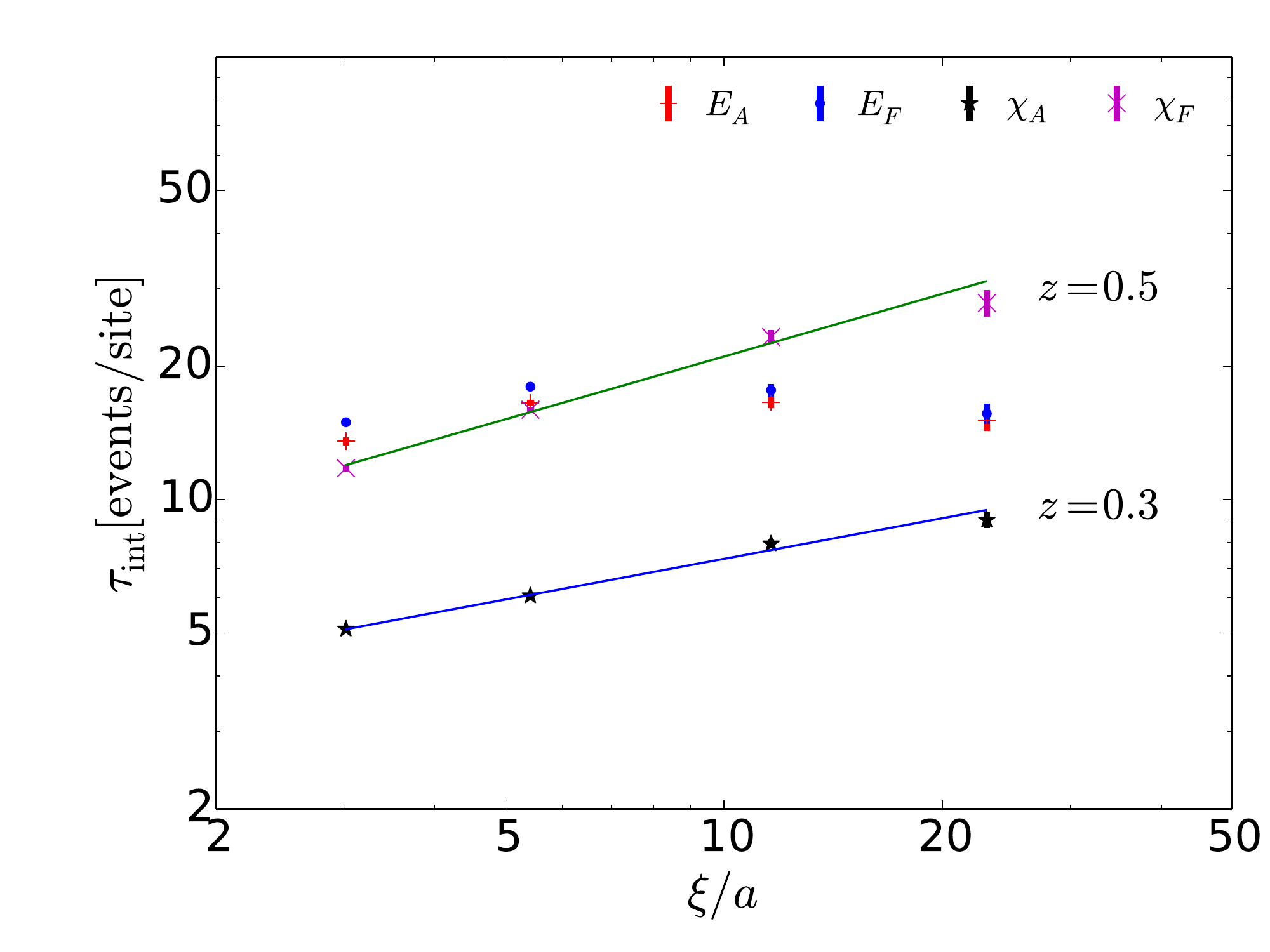}
  \end{center}
  \caption{Scaling of the integrated autocorrelation times with the correlation length. We observe dynamical critical exponents $z$ well below $1$ in all cases.\label{f3}}
\end{figure}

In Fig.~\ref{f3} we plot the minimal integrated autocorrelation  against
the correlation length in Fig.~\ref{f3}. The scaling is compatible with
$z < 1$ for all observables that we have measured.

\section{Conclusions and Outlook}
We applied the event-chain Monte 
Carlo algorithm~\cite{PhysRevE.80.056704,0295-5075-112-2-20003,PhysRevE.92.063306,KrauthPRL} to asymptotically free models in two dimensions. 
In the literature, these models are considered as toy models of lattice QCD 
that allow for 
testing of algorithmic and theoretical ideas in a simplified setting.
In a preliminary study, we applied the  event-chain Monte Carlo to free field
theory on a square and a simple cubic lattice. We find that critical slowing 
down is eliminated. This is quite astonishing for an algorithm, where local 
changes of the field variables are performed.  
Next we studied the two-dimensional XY-model. We do not discuss our results
in detail. They corroborate the findings of \cite{Lei:2018vis}: At low
temperatures, in the spin wave phase, slowing down is eliminated, which is
consistent with our free field theory result. On the other hand, in the high 
temperature phase, the physics is governed by vortices. Here we find
slowing down with $z \approx 2$, even though the amplitudes of 
the integrated autocorrelation times are considerably reduced compared with 
the local Metropolis algorithm. One should note however that for this model the 
overrelaxation algorithm \cite{Apo91}  gives $z \approx 1$ in both phases 
and the cluster algorithm eliminates slowing down completely 
\cite{UWolff89,Wolff:1988kw}.
In order to apply the event-chain Monte algorithm to the $O(3)$-invariant 
nonlinear $\sigma$-model and the $SU(3) \times SU(3)$ chiral model,
we use an embedding of the 
XY model.  In the case of the $O(3)$-invariant model this simply means that
one picks out two of the three components of the field. During one sequence
of updates only these two components are updated. For the next sequence 
a new pair of components is taken. In the case of the 
$SU(3) \times SU(3)$ principal chiral 
model the embedding is a bit more complicated. We follow ref. 
\cite{Mendesaetal96} who used such an embedding in the context of the 
multigrid algorithm. For details see section \ref{principalchiral}.

We find for both the $O(3)$-invariant nonlinear $\sigma$-model and the 
$SU(3) \times SU(3)$ chiral model that critical slowing down 
is essentially eliminated.  While this is certainly encouraging with regard 
to the application to lattice QCD, open questions remain. How should 
a coherent embedding of $U(1)$ variables in the gauge theory look like?
Will the event-chain Monte Carlo algorithm speed up the decorrelation 
of topological object that are present in QCD? 

\appendix

\section{Proof of stability \label{a1}}
We have to show that 
the infinitesimal update by $\epsilon$ preserves the target distribution
\begin{equation}
\pi(X)= \frac{1}{2 V Z} \exp(-S[\phi]) \;\;, 
\end{equation}
where $Z$ is the partition function of the lattice model and the factors
$2$ and $V$ give the number of possible values of $\sigma$ and $\tilde x$.
Let us write the update probabilities defined in section \ref{eventalg}
as $T(Y\leftarrow X)$. Then stability means 
\begin{equation}
\label{stability}
  \pi(Y) = \sum_X T(Y\leftarrow X) \pi(X) \;\;,
\end{equation}
where $\sum_X$ is a short hand for the integral over the field $\phi$ and 
the sums over $\tilde x$ and $\sigma$. 
As a first step, we  list the configurations $X$ that can end up  
in a given $Y=(\phi, \sigma, \tilde x)$, after an update step:
\begin{itemize}
\item $X_{0}$ given by $\phi$ the same as for $Y$, $\sigma$ is replaced by 
$-\sigma$ and the position of the walker $\tilde x$ remains the same.

\item $X_i$ given by $\phi$ is the same as for $Y$, $\sigma$ is the same as 
for $Y$ and $\tilde x = y_i$,
meaning that the walker is hopping to its interaction 
partner $y_i$.

  \item  $X_{n+1}$ given by $\phi_{\tilde x} -\sigma \epsilon$,  
  $\phi_x$ with $x\ne \tilde x$ are 
        the same as for $Y$, $\sigma$ and $\tilde x$ keep their value.
\end{itemize}
Hence eq.~(\ref{stability}) can be written as
\begin{equation}
\label{stability2}
  \pi(Y) = \sum_{i=0}^{n+1}  T(Y\leftarrow X_i) \pi(X_i) \;\;.
\end{equation}
Note that the probability density of $X_i$ for $i \le n$ is the same as 
that of $Y$, since the field $\phi$ is the same.
The probability density of $X_{n+1}$ is 
\begin{eqnarray}
    \pi(X_{n+1})  &=& \pi(Y) \;
   \exp\left[-\sum_i [s_{i,x}(\phi_{\tilde x} -\sigma \epsilon)-s_{i,x}
   (\phi_{\tilde x}) ]\right] \nonumber \\
&=& \pi(Y) \; \left[1 + 
\sigma \sum_{i=0}^n
     \frac{\partial s_{i,\tilde x}}{\partial \phi_{\tilde x}}  \epsilon
 + O(\epsilon^2) \right] \;\;.
\label{BX1}
\end{eqnarray}
Now let us work out the transition probabilities:
  \begin{equation}
    T(Y \leftarrow X_i)= 1 - p_i = \mbox{max}\left[0, - \sigma 
  \frac{\partial s_{i,\tilde x}}{\partial \phi_{\tilde x}} \epsilon 
  \right] + O(\epsilon^2)  \;\;\;\; \mbox{for} \;\;\; i \le n \;\;.
  \end{equation}
Note that for $i=0$ the sign of $\sigma$ changes, while for $1 \le i \le n$
the hopping of the walker along with eq.~(\ref{symab}) produces a minus sign.
Finally
  \begin{eqnarray}
    T(Y \leftarrow X_{n+1}) = \prod_{i=0}^n  p_i   
   &=& \prod_{i=0}^n \mbox{min}
    \left[1,1-\sigma 
     \frac{\partial s_{i,\tilde x}}{\partial \phi_{\tilde x}} \epsilon
\right] + O(\epsilon^2) \nonumber \\
&=& 1 + \sum_{i=0}^n \mbox{min} \left[0, - \sigma
 \frac{\partial s_{i,\tilde x}}{\partial \phi_{\tilde x}} \epsilon \right]  + O(\epsilon^2) \,.
  \end{eqnarray}

Now we can put things together:
\begin{eqnarray}
&&\sum_{i=0}^{n}  T(Y\leftarrow X_i) \pi(X_i) + 
T(Y\leftarrow X_{n+1}) \pi(X_{n+1})  \nonumber \\
&=& \pi(Y) 
\left( \sum_{i=0}^{n} 
 \mbox{max}\left[0,- \sigma
  \frac{\partial s_{i,\tilde x}}{\partial \phi_{\tilde x}} \epsilon \right]
+ O(\epsilon^2) \right) \nonumber \\
&+& \pi(Y)
\left (1 +
\sigma \sum_{i=0}^n
 \frac{\partial s_{i,\tilde x}}{\partial \phi_{\tilde x}}  \epsilon 
+ O(\epsilon^2) \right)
  \left(1 + \sum_{i=0}^n \mbox{min} \left[0, - \sigma
 \frac{\partial s_{i,\tilde x}}{\partial \phi_{\tilde x}} \epsilon \right] 
+ O(\epsilon^2)
\right)  \nonumber \\
&=& \pi(Y) \left[1 + O(\epsilon^2) \right] \;\;.
\end{eqnarray}
Contributions $O(\epsilon)$ exactly cancel each other.


\begin{thebibliography}{10}

\bibitem{PhysRevE.80.056704}
E. P. Bernard, W. Krauth, and D. B. Wilson, {\sl Event-chain Monte Carlo
  algorithms for hard-sphere systems}, Phys.\ Rev.\ E {\bf 80}, 056704
 (2009).

\bibitem{0295-5075-112-2-20003}
M. Michel, J. Mayer, and W. Krauth, {\sl Event-chain Monte Carlo for classical
  continuous spin models}, EPL (Europhysics Letters) {\bf 112}, 
  20003 (2015).

\bibitem{PhysRevE.92.063306}
Y. Nishikawa, M. Michel, W. Krauth, and K. Hukushima, 
{\sl Event-chain algorithm
 for the Heisenberg model: Evidence for $z\ensuremath{\simeq}1$ dynamic
  scaling}, Phys.\ Rev.\ E {\bf 92}, 063306 (2015).


\bibitem{KrauthPRL}
S. Kapfer and W. Krauth, {\sl Irreversible Local Markov Chains with Rapid
  Convergence towards Equilibium}, Phys.\ Rev.\ Lett.\ {\bf 119}, 
  240603, (2017).

\bibitem{Swendsen:1987ce}
R. H. Swendsen and J.-S. Wang, 
{\sl Nonuniversal critical dynamics in Monte Carlo
 simulations}, Phys.\ Rev.\ Lett.\ {\bf 58}, 86 (1987).

\bibitem{Wolff:1988uh}
U. Wolff, {\sl Collective Monte Carlo Updating for Spin Systems},
  Phys.\ Rev.\ Lett.\ {\bf 62}, 361 (1989).

\bibitem{Goodman:1989jw}
J. Goodman and A. D. Sokal, {\sl Multigrid Monte Carlo method. Conceptual
  foundations}, Phys.\ Rev.\ D {\bf 40}, 2035 (1989).

\bibitem{Brower:1989mt}
R. C. Brower and P. Tamayo, {\sl Embedded dynamics for $\phi^4$ theory},
  Phys.\ Rev.\ Lett.\ {\bf 62}, 1087 (1989).

\bibitem{Mendesaetal96}
T. Mendes, A. Pelissetto, and A. D. Sokal, {\sl Multi-grid Monte Carlo via XY
  embedding. General theory and two-dimensional $O(N)$-symmetric non-linear
  $\sigma$-models}, Nucl.\ Phys.\ B {\bf 477}, 203 (1996).

\bibitem{HaPi96}
M. Hasenbusch and K. Pinn, {\sl Computing the roughening transition of Ising 
and solid-on-solid models by BCSOS model matching},
J.\ Phys.\ A {\bf 30}, 63 (1997).

\bibitem{Ha08}
M. Hasenbusch, {\sl The Binder cumulant at the Kosterlitz-Thouless transition},
  J.\ Stat.\ Mech.: Theor. Exp. (2008) P08003. 

\bibitem{Balog:1999ww}
J. Balog, M. Niedermaier, F. Niedermayer, A. Patrascioiu, E. Seiler, and
  P. Weisz, {\sl Comparison of the O(3) bootstrap $\sigma$ model with the 
lattice
  regularization at low-energies}, Phys.\ Rev.\ D {\bf 60}, 094508
  (1999).

\bibitem{UWolff89}
U.~Wolff, {\sl Asymptotic freedom and mass generation in the O(3) nonlinear
  $\sigma$-model}, Nucl.\ Phys.\ B {\bf 334}, 581 (1990).

\bibitem{Apo91}
J. Apostolakis, C. F. Baillie, and G. C. Fox, {\sl Investigation of the
 two-dimensional O(3) model using the overrelaxation algorithm}, 
 Phys.\ Rev.\ D {\bf 43}, 2687 (1991).

\bibitem{UlliWurm}
U. Wolff, {\sl Simulating the all-order strong coupling expansion III: $O(N)$
  sigma/loop models}, Nucl.\ Phys.\ B {\bf 824}, 254 (2010).
\newblock [Erratum: Nucl.\ Phys.\ B {\bf 834}, 395 (2010)].

\bibitem{Mana:1996pk}
G. Mana, A. Pelissetto, and A. D. Sokal, {\sl Multigrid Monte Carlo via XY
 embedding. II. Two-dimensional SU(3) principal chiral model}, 
Phys.\ Rev.\ D {\bf 55}, 3674 (1997).

\bibitem{Lei:2018vis}
Z. Lei and W. Krauth, {\sl Irreversible Markov chains in spin models: 
Topological excitations}, EPL (Europhysics Letters) {\bf 121}, 10008 (2018).

\bibitem{balogetal99}
J. Balog, M. Niedermaier, F. Niedermayer, A. Patrascioiu, E. Seiler, and
 P. Weisz, {\sl Comparison of the O(3) bootstrap $\sigma$ model with the lattice
  regularization at low-energies}, Phys.\ Rev.\ D {\bf 60}, 094508
  (1999).

\bibitem{Hasenbusch:1991aj}
M. Hasenbusch and S. Meyer, {\sl Multigrid acceleration for asymptotically 
free theories},
Phys.\ Rev.\ Lett.\ {\bf 68}, 435 (1992).

\bibitem{Wolff:1988kw}
U. Wolff, {\sl Collective Monte Carlo updating in a high precision study of
 the
  x-y Model}, Nucl.\ Phys.\ B {\bf 322}, 759 (1989).

\end{thebibliography}
\end{document}